\documentclass[twocolumn,epjc3]{svjour3}  
\smartqed  
\RequirePackage{graphicx}
\RequirePackage[colorlinks,citecolor=blue,urlcolor=blue,linkcolor=blue]{hyperref}
\RequirePackage{xspace}
\RequirePackage{amsmath}
\usepackage{lineno}
\newcommand{\gerda}   {\textsc{Gerda}\xspace}
\newcommand{\PI}      {Phase~I\xspace}
\newcommand{\PII}     {Phase~II\xspace}

\newcommand{\Beta}    {\ensuremath{\beta}\xspace}
\newcommand{\nnbb}    {\ensuremath{2\nu\beta\beta}\xspace}
\newcommand{\onbb}    {\ensuremath{0\nu\beta\beta}\xspace}
\newcommand{\onecec}  {0\ensuremath{\nu}ECEC\xspace}
\newcommand{\nnecec}  {2\ensuremath{\nu}ECEC\xspace}
\newcommand{\bege}    {\ensuremath{^{enr}}BEGe\xspace}
\newcommand{\coax}    {\ensuremath{^{enr}}Coax\xspace}
\newcommand{\nat}     {\ensuremath{^{nat}}Coax\xspace}
\newcommand{\ic}      {\ensuremath{^{enr}}IC\xspace}
\newcommand{\Ge}      {\ensuremath{^{76}}Ge\xspace}
\newcommand{\Ar}      {\ensuremath{^{36}}Ar\xspace}

\newcommand{\qbb}     {\ensuremath{Q_{\beta\beta}}\xspace}
\newcommand{\geant}   {\textsc{Geant4}\xspace}
\newcommand{\mage}    {\textsc{MaGe}\xspace}
\newcommand{\invHL}   {\ensuremath{T_{1/2}^{-1}}\xspace}
\newcommand{\HL}   {\ensuremath{T_{1/2}}\xspace}
\newcommand{\cls}     {\ensuremath{\text{CL}_\text{s}}\xspace}
\newcommand{\pvalue}  {\ensuremath{p\text{-value}}\xspace}
\journalname{Eur. Phys. J. C}
\begin{document}

\title{An improved limit on the neutrinoless double-electron capture of \Ar with GERDA}


\author{
The \mbox{\protect{\sc{Gerda}}} collaboration\thanksref{corrauthor}
\and  \\[4mm]
M.~Agostini\thanksref{UCL} \and
A.~Alexander\thanksref{UCL} \and
G.R.~Araujo\thanksref{UZH} \and
A.M.~Bakalyarov\thanksref{KU} \and
M.~Balata\thanksref{ALNGS} \and
I.~Barabanov\thanksref{INRM} \and
L.~Baudis\thanksref{UZH} \and
C.~Bauer\thanksref{HD} \and
S.~Belogurov\thanksref{ITEP,INRM,alsoMEPHI} \and
A.~Bettini\thanksref{PDUNI,PDINFN} \and
L.~Bezrukov\thanksref{INRM} \and
V.~Biancacci\thanksref{LNGSGSSI} \and
E.~Bossio\thanksref{TUM} \and
V.~Bothe\thanksref{HD} \and
V.~Brudanin\thanksref{JINR} \and
R.~Brugnera\thanksref{PDUNI,PDINFN} \and
A.~Caldwell\thanksref{MPIP} \and
C.~Cattadori\thanksref{MIBINFN} \and
A.~Chernogorov\thanksref{ITEP,KU} \and
T.~Comellato\thanksref{TUM}
% ...
}
\authorrunning{the \textsc{Gerda} collaboration}
\thankstext{corrauthor}{
  \emph{correspondence}  gerda-eb@mpi-hd.mpg.de}
\thankstext{alsoMEPHI}{\emph{also at:} NRNU MEPhI, Moscow, Russia}
\thankstext{nowDuke}{\emph{now at:} Duke University, Durham, NC USA}
\thankstext{alsoLev}{\emph{also at:} Moscow Inst. of Physics and Technology,
  Russia}
\thankstext{alsoDubna}{\emph{also at:} Dubna State University, Dubna, Russia}
\thankstext{alsoRM3}{\emph{also at:} INFN Roma Tre, Rome, Italy}
\institute{
INFN Laboratori Nazionali del Gran Sasso, Assergi, Italy\label{ALNGS} \and
INFN Laboratori Nazionali del Gran Sasso and Gran Sasso Science Institute, Assergi, Italy\label{LNGSGSSI}
% ...
\and
Joint Institute for Nuclear Research, Dubna, Russia\label{JINR}
% ...
\and
Max-Planck-Institut f{\"u}r Kernphysik, Heidelberg, Germany\label{HD} \and
Department of Physics and Astronomy, University College London, London, UK\label{UCL} \and
INFN Milano Bicocca, Milan, Italy\label{MIBINFN}
% ...
\and
Institute for Nuclear Research of the Russian Academy of Sciences, Moscow, Russia\label{INRM} \and
Institute for Theoretical and Experimental Physics, NRC ``Kurchatov Institute'', Moscow, Russia\label{ITEP} \and
National Research Centre ``Kurchatov Institute'', Moscow, Russia\label{KU} \and
Max-Planck-Institut f{\"ur} Physik, Munich, Germany\label{MPIP} \and
Physik Department, Technische  Universit{\"a}t M{\"u}nchen, Germany\label{TUM} \and
Dipartimento di Fisica e Astronomia, Universit{\`a} degli Studi di 
Padova, Padua, Italy\label{PDUNI} \and
INFN  Padova, Padua, Italy\label{PDINFN}
% ...
\and
Physik-Institut, Universit{\"a}t Z{\"u}rich, Z{u}rich, Switzerland\label{UZH}
}
%


\maketitle

\begin{abstract}
The GERmanium Detector Array (\gerda) experiment operated enriched high-purity germanium detectors in a liquid argon cryostat, which contains 0.33\% of \Ar, a candidate isotope for the two-neutrino double-electron capture (\nnecec) and therefore for the neutrinoless double-electron capture (\onecec). If detected, this process  would give evidence of lepton number violation and the Majorana nature of neutrinos. In the radiative \onecec of \Ar, a monochromatic photon is emitted with an energy of 429.88\,keV, which may be detected by the \gerda germanium detectors. We searched for the \Ar \onecec with \gerda data, with a total live time of 4.34\,yr (3.08\,yr accumulated during \gerda \PII and 1.26\,yr during \gerda \PI). No signal was found and a 90\% C.L. lower limit on the half-life of this process was established $T_{1/2} > 1.5 \cdot 10^{22}$\,yr. 

\end{abstract}

\section{Introduction}\label{sec:intro}

The simultaneous capture of two bound atomic electrons followed by the emission of two neutrinos plus X-rays or Auger electrons, known as two-neutrino double-electron 
capture (\nnecec), is a nuclear process allowed in the Standard Model. Compared to the two-neutrino double-beta (\nnbb) decay, the simultaneous emission of two electrons and 
two anti-neutrinos, \nnecec processes have lower probabilities due to the smaller phase space, therefore experimentally, they are much more challenging to observe. 
The first direct observation of \nnecec was made only in 2018 by the XENON1T experiment with $^{124}$Xe~\cite{XenonCollaboration2019}. 
Previously, indications of \nnecec were found in geochemical measurements with $^{130}$Ba and $^{132}$Ba~\cite{Meshik2001} and in a large proportional counter experiment 
with $^{78}$Kr~\cite{Ratk2017}.

The lepton number violating counterpart of \nnecec, the neutrinoless double-electron capture (\onecec), in which no neutrinos are emitted, is also predicted~\cite{Winter1955}. This process must be accompanied by the emission of at least another particle to ensure energy and momentum conservation.
Different modes can be considered in which \onecec is associated with the emission of different particles like $e^+e^-$ pairs, one or two photons, or one internal conversion electron~\cite{Doi1993,Blaum2020}. 
In analogy with the neutrinoless double-beta (\onbb) decay, the \onecec violates the lepton number symmetry by two units and implies that neutrinos have a Majorana mass component~\cite{Georgi1981}. Although the sensitivity of \onecec processes to the Majorana neutrino mass is estimated to be many orders of magnitude lower than that of the \onbb decay, the interest in \onecec is theoretically motivated by the possibility of resonant enhancement when the parent nucleus and an excited state of the daughter nucleus are energetically degenerate~\cite{Winter1955,Blaum2020,Georgi1981,Voloshin1982,Bernabeu1983}. In this case, the half-life of \onecec processes becomes comparable to that of \onbb decays. Experimental searches for \onecec have been performed by double-\Beta decay experiments, even though with less sensitivity compared to the search for \onbb decay~\cite{Blaum2020}. 

The GERmanium Detector Array (\gerda) experiment, whose main goal was to search for the \onbb decay of \Ge~\cite{GERDA:2012qwd,GERDA:2020xhi}, operated enriched high purity germanium detectors in a liquid argon (LAr) cryostat, which naturally contains the \Ar isotope with an isotopic abundance of 0.33\%. \Ar can undergo \nnecec to the ground state of $^{36}$S~\cite{Tretyak1995}. The corresponding lepton number violating process, \onecec, may occur via the simplest radiative mode\footnote{Given the available energy of the process, the internal conversion mode would also be allowed for \Ar. Nevertheless, the latter is strongly suppressed due to argon's low atomic number and the relatively high $\gamma$ energy~\cite{Merle2009}.}
\begin{equation}
    \text{\Ar} \rightarrow \, ^{36}\text{S} + \gamma \;  + \text{X}_\text{K} \; + \text{X}_\text{L} \;.
\end{equation}
The \Ar nucleus captures one electron each from its K- and L-shells and turns into $^{36}$S. Two X-rays are emitted, with energies E$_\text{K} = 2.47$\,keV, and E$_\text{L} = 0.23$\,keV, corresponding to the capture of the electrons from the K- and the L-shell, respectively. Given the available energy of the decay Q$_\text{ECEC} = (432.58 \pm 0.19)$\,keV~\cite{Wang2012}, the corresponding energy for the $\gamma$ ray is E$_\gamma = \text{Q}_\text{ECEC}-\text{E}_\text{K} -\text{E}_\text{L} = (429.88 \pm 0.19)$\,keV.
Resonance enhancement of the process is not possible for \Ar~\cite{Blaum2020}. In the light neutrino exchange scenario, assuming a Majorana mass of 0.1\,eV, the half-life of \Ar \onecec is predicted in the order of $10^{40}$\,yr, with calculations based on the quasiparticle random-phase approximation (QRPA)~\cite{Merle2009}. 
Experimental searches for \onecec of \Ar have been performed since the early stages of the \gerda experiment~\cite{Chkvorets2008}. The most stringent limit to date on the \Ar \onecec half-life is T$_{1/2} > 3.6 \times 10^{21}$\,yr (90\% C.I.), established in \PI of the \gerda experiment~\cite{Agostini2016}. More recently, this process has been searched with the DEAP detector~\cite{Dunford2018}, although with less sensitivity than \gerda \PI. 

In this paper, we report on the search for the 429.88\,keV $\gamma$ line from the \Ar \onecec with the whole \gerda data, accumulated for a total live time of 3.08\,yr during \gerda \PII and 1.26\,yr during \gerda \PI.

\section{The GERDA experiment}\label{sec:gerda}

The \gerda experiment was located at the Laboratori Nazionali del Gran Sasso (LNGS) of INFN, in Italy~\cite{GERDA:2012qwd,GERDA:2017ihb,Agostini:2019hzm}, where a rock overburden of 3500\;m water equivalent reduces the flux of cosmic muons by six orders of magnitude~\cite{GERDA:2012qwd}. High-purity germanium (HPGe) detectors, isotopically enriched in \Ge, were operated inside a 64\;m$^3$ LAr cryostat~\cite{Knopfle:2022fso}. 
In the second phase of the experiment, 10 coaxial (including 3 detectors with natural isotopic abundance) and 30 Broad Energy Germanium (BEGe) detectors were used~\cite{GERDA:2017ihb}. After an upgrade in May 2018, the three natural coaxial detectors were removed, and 5 additional inverted coaxial (IC) detectors were installed~\cite{GERDA:2020xhi}. Detectors were mounted on 7 strings, and each string was placed inside a nylon cylinder to limit the collection of radioactive potassium ions on the detector surfaces~\cite{Lubashevskiy:2017lmf}. 
The LAr volume around the detectors was instrumented with a curtain of wavelength-shifting fibers connected to silicon photo-multipliers (SiPM) and 16 cryogenic photo-multiplier tubes (PMTs) to detect scintillation light in the LAr~\cite{Janicsko-Csathy:2010uif,GERDA:2017ihb}. During the upgrade, the geometrical coverage of the fibers was improved, more SiPM channels were added, and their radiopurity increased~\cite{GERDA:2020xhi}.
The cryostat was surrounded by a water tank containing 590\;m$^3$ of pure water, equipped with PMTs to detect the Cherenkov light of residual cosmic muons reaching the detector site. The instrumented water tank formed, together with scintillator panels on the top of the experiment, the muon veto system~\cite{Freund:2016fhz}.  

\section{Data selection}\label{sec:exp-data}

The \gerda \PII data taking started in December 2015; it was shortly interrupted in the Summer of 2018 for the upgrade of the setup and lasted until November 2019. The total collected data used to search for the 429.88\,keV $\gamma$ line from the \onecec of \Ar corresponds to a live time of 3.08\,yr, divided into 1.91\,yr before the upgrade and 1.17\,yr after the upgrade. Due to the different detector properties, \emph{e.g.} energy resolution and efficiency, and the changes in the detector configuration during the upgrade, data were split into 5 data sets, namely pre-upgrade \bege, pre-upgrade \coax, post-upgrade \bege, post-upgrade \coax, and post-upgrade \ic. The \nat detectors were excluded from the analysis since they have a low duty factor due to their unstable operation in \gerda \PII and made up a minimal amount of the exposure. 

Data have been processed following the procedures and digital signal processing algorithms described in~\cite{Agostini:2011mh}. The energy of an event is reconstructed using a zero-area-cusp filter~\cite{GERDA:2015rik}. 
Events must pass several quality cuts based on the flatness of the baseline, polarity, and time structure of the pulse to reject non-physical events. The acceptance efficiency of physical events by quality cuts is larger than 99.9\%~\cite{GERDA:2020xhi}. Events preceded by a trigger in the muon-veto system within 10\,$\mu$s are also discarded, with negligible induced dead time ($<$0.01\%)~\cite{GERDA:2020xhi}. 

\begin{figure}
  \includegraphics[width=\columnwidth]{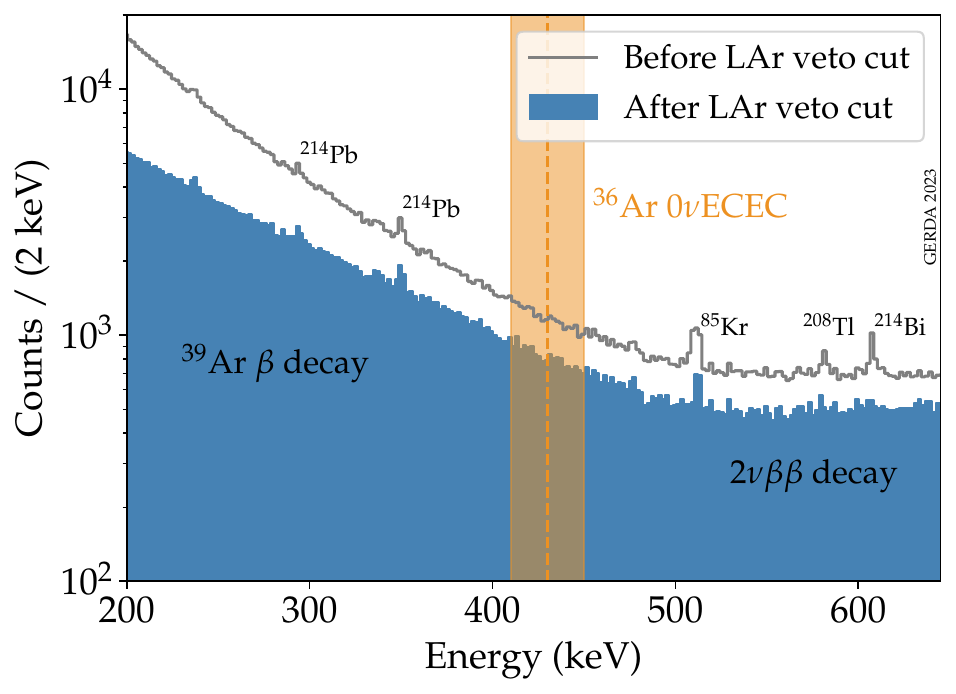}
\caption{Energy distribution of the low energy \gerda \PII data before and after LAr veto cut. The left part of the spectrum is dominated by the $^{39}$Ar \Beta decay with an endpoint at 565\,keV. On the right side, the \nnbb decay dominates. Some known $\gamma$ lines are visible and labeled. The orange dotted line indicates the energy at which the \Ar \onecec is expected, and the orange band indicates the energy region used in the analysis. }
\label{fig:ar36_lowenergy_spectrum}      
\end{figure}

The experimental signature used to search for \Ar \onecec in the \gerda data corresponds to the full energy deposition of the $\gamma$ ray in one germanium detector. Neglecting the energy deposition of the two X-rays, no coincident energy deposition is expected, neither in the other germanium detectors nor the LAr. Consequently, the detector anti-coincidence cut and the LAr veto cut were also applied. The energy of the two X-rays is low enough that, even if they reached the germanium detector surface, they could not penetrate the 1--2\,mm dead layer and, therefore, not be detected by the germanium detector. Nevertheless, since they deposit their energy in the LAr, they could be seen by the LAr instrumentation and trigger the LAr veto. The corresponding event would escape the data selection. This effect is considered in the total detection efficiency, as will be explained in section~\ref{sec:efficiency}.
The LAr veto cut reduces the background in the region of interest of this analysis by a factor of $\sim 2$, as can be seen in Figure~\ref{fig:ar36_lowenergy_spectrum}. In this energy region, $^{39}$Ar \Beta decay dominates up to the endpoint at 565\,keV, while \nnbb decay is the second dominant contribution.  
The Pulse Shape Discrimination (PSD) cut, successfully employed in the search for \onbb decay~\cite{Agostini2022}, is unsuitable for this analysis and, therefore, not used. In fact, $\gamma$ rays mostly result in multiple separated energy depositions in the germanium detector, {\it i.e.} multi-site events, in contrast to the single-site events produced in the \onbb decay. In addition, the performances of the PSD cut at the energy of interest of this analysis are poorly known.
Consequently, part of the data excluded in the \onbb decay analysis from \bege and \ic data sets because of the PSD cut was instead included here.  

We combine the analysis of \gerda \PII data with that of \gerda \PI data reported in~\cite{Agostini2016}. The \gerda \PI data taking started in November 2011 and lasted until May 2013. The total collected data used for searching for \onecec of \Ar corresponded to a live time of 1.26\,yr and was divided into three data sets, namely \coax, \bege, and \nat. More details on the data processing and selection of these three data sets can be found in~\cite{Agostini2016}. It has to be noticed that the instrumentation of the LAr volume is a unique feature of \gerda \PII and that no LAr veto cut was available in \gerda \PI. 

\section{Energy resolution and energy scale}\label{sec:low-energy-cal}

The energy calibration of the \gerda germanium detectors was performed during dedicated weekly calibration runs in which the germanium detectors were exposed to three $^{228}$Th sources~\cite{GERDA:2021pcs}. All calibration data were combined as detailed in~\cite{GERDA:2021pcs} to determine the energy scale and resolution throughout the experiment. 

\begin{figure}
  \includegraphics[width=\columnwidth]{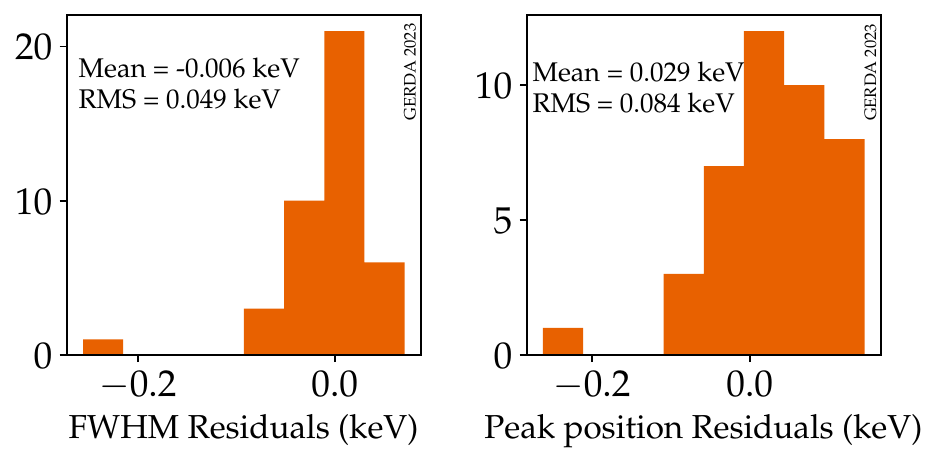}
\caption{(Left) Distribution of the energy resolution (FWHM) residuals for the 583\,keV calibration peak. (Right) Distribution of the peak position residuals for the same calibration peak. The mean and the RMS of the two distributions are indicated.}
\label{fig:residuals-cal}      
\end{figure}

This work uses the effective resolution curves calculated for the five analysis data sets~\cite{Ransom:2021fcg}. The resolution curves are evaluated at the \Ar \onecec $\gamma$ energy of 429.88\,keV. 
The energy resolution in full width at half maximum (FWHM) and their uncertainties are summarized in table~\ref{tab:res-eff}. The uncertainty on the FWHM is calculated assuming the same relative uncertainty as for the FWHM at the \qbb of the \Ge \onbb decay (\qbb = 2039\,keV). This was calculated in~\cite{GERDA:2021pcs} as exposure-weighted standard deviation. 
The picture might be different at low energy, and the results obtained for the \onbb decay peak at 2039\,keV might not be valid for the \onecec peak at 429.88\,keV. In fact, the lowest energy peak used to determine the resolution curves above is the 583\,keV $^{208}$Tl peak, above the energy region of interest in this analysis. To cross-check the energy resolution at the energy of interest, we use the results of the special low-energy calibration performed at the end of the \gerda data taking. This calibration run aimed to study the energy scale and stability at low energy. The energy threshold was set to 100\,keV (while it was 400\,keV during regular calibration runs), allowing to extend the energy range in which the resolution curve is calculated to about 238\,keV, the energy of the first $^{212}$Pb $\gamma$ peak usable for the calibration. We use the peak at 583\,keV as a proxy for the \onecec peak, being the closest in energy. We should note that also the topology of the events for the two peaks is the same. In both cases, it is a full energy deposition of the $\gamma$ energy in one germanium detector, with the $\gamma$ ray starting in the surrounding of the detector array. We calculate the residuals on the FWHM as the difference between the FWHM extracted in the special low-energy calibration and the value obtained evaluating the resolution curves above at 583\,keV. The residuals for each detector are shown in a histogram at the left-handed side of figure~\ref{fig:residuals-cal}. We find no systematic deviation of the FWHM at this energy compared to the resolution curves. The RMS of the residuals is 0.049\,keV, with only one detector with a larger residual of -0.2\,keV.\footnote{This is a \coax detector, so the result is compatible with the larger FWHM uncertainty of the post-upgrade \coax data set.}

The monitoring of the energy scale for the \onbb decay search was performed using the single escape peak of $^{208}$Tl at 2103\,keV, which is typically used as a proxy for the \onbb decay peak at \qbb. The residuals between the peak position after energy calibration and the nominal energy value were evaluated over time, giving a mean energy bias of -0.07\,keV with an average uncertainty of 0.17\,keV~\cite{GERDA:2021pcs}. To cross-check the energy bias at the energy of interest, we use the results of the special low energy calibration run and the 583\,keV peak as a proxy for the \onecec peak again. We calculate the residuals on the peak position as the difference between the nominal energy value and the energy value extracted from the special low-energy calibration. The residuals for each detector are shown in a histogram at the right-handed side of figure~\ref{fig:residuals-cal}. We find a mean energy bias of 0.03\,keV with a RMS among detectors of 0.084\,keV. This is below the estimated bias uncertainty of 0.17\,keV for the \onbb decay peak at \qbb. It should be noted that these biases are well below the binning of 1\,keV used in the analysis. The effect is therefore expected to be marginal. In this work, we adopt a mean energy bias of 0\,keV with an uncertainty of 0.1\,keV for all the five analysis data sets. 

\begin{table}
\caption{Energy resolution (FWHM) and $\gamma$ detection efficiency (multiplied by the simulated mass of LAr) for the analysis data sets. The values for the \PI data sets are taken from~\cite{Agostini2016}.}
\label{tab:res-eff}
\begin{tabular}{lll}
\hline\noalign{\smallskip}
Data set & FWHM (keV) & $\varepsilon_\gamma \cdot m_{LAr}$ (kg) \\
\noalign{\smallskip}\hline\noalign{\smallskip}
\multicolumn{2}{l}{\PII pre-upgrade} \vspace{0.6mm} \\
\bege & $2.2 \pm 0.2$ & $2.04 \pm 0.06$ \\
\coax & $2.7 \pm 0.2$ & $1.72 \pm 0.06$ \vspace{0.6mm} \\
\multicolumn{2}{l}{\PII post-upgrade} \vspace{0.6mm} \\
\bege & $1.74 \pm 0.09$ & $2.51 \pm 0.07$ \\
\coax & $3.1 \pm 1.3$ & $1.29 \pm 0.06$ \\
\ic & $1.72 \pm 0.07$ & $0.841 \pm 0.006$ \vspace{0.6mm} \\
\multicolumn{2}{l}{\PI} \vspace{0.6mm} \\
\coax & $3.72 \pm 0.05$ & $1.79 \pm 0.18$ \\
\bege & $2.01 \pm 0.10$ & $0.281 \pm 0.018$ \\
\nat & $4.08 \pm 0.20$ & $0.739 \pm 0.073$ \\
\noalign{\smallskip}\hline
\end{tabular}
\end{table}

\section{Detection efficiency}\label{sec:efficiency}

\begin{figure}
    \centering
    \includegraphics[width=\columnwidth]{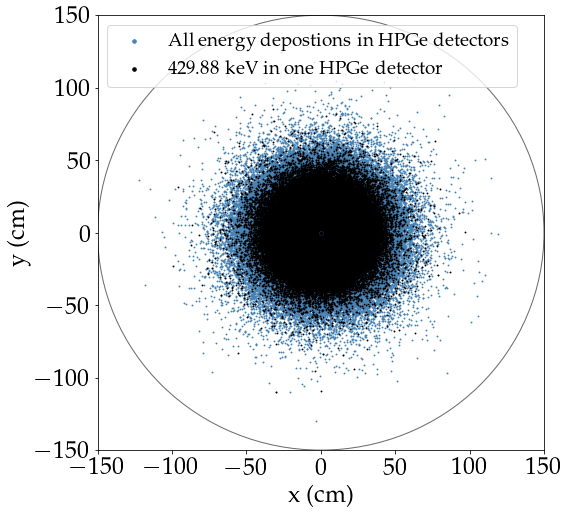}
    \caption{Projected distribution of vertices from which the simulated $\gamma$ rays originate. $\gamma$ rays with an energy of 429.88\,keV are simulated uniformly in the cylindrical volume. Only those originating from the blue vertices deposit some energy in the germanium detectors, while only those originating from the back vertices deposit the entire energy in one germanium detector, thus contributing to the $\gamma$ detection efficiency. }
    \label{fig:eff_sim}
\end{figure}

The $\gamma$ detection efficiency is defined as the probability that a 429.88\,keV $\gamma$ ray entirely deposits its energy inside a single germanium detector. This was determined via Monte Carlo simulations with the \geant-based \mage framework~\cite{Agostinelli2003,Boswell2011}. In total, $10^{10}\; \gamma$ rays with an energy of 429.88\,keV were generated in a cylindrical volume of LAr, with a radius of 1.5\,m and a height of 2.5\,m, around the detector array. This corresponds to a net volume of LAr, after taking into account the volume occupied by the germanium detectors and structural materials, of 17.657\,m$^3$. The corresponding LAr mass, given the LAr density of 1385\,kg/m$^3$, is 24459\,kg. 
The contribution from $\gamma$ rays originating from outside this volume to the detection efficiency is negligible, as shown in figure~\ref{fig:eff_sim}. The projected distribution of vertices from which the simulated $\gamma$ rays originate is shown in blue for all the events resulting in an energy deposition in the germanium detectors and black for the events resulting in the deposition of the entire 429.88\,keV $\gamma$ energy in one germanium detector.
Only the last contribute to the $\gamma$ detection efficiency, defined for each data set as the ratio between the number of events in which the full energy is deposited in one germanium detector in the specific data set and the number of initially simulated events. The number of simulated events is high enough that the statistical uncertainties on these quantities are negligible. Detector active volume and the status of each detector over the whole data taking are considered in the simulation, as detailed in~\cite{GERDA:2019cav}. The dominant systematic uncertainty on the $\gamma$ detection efficiency comes from the detector active volume uncertainty. This is estimated by varying the detector dead layer in the simulation by $\pm 1\,\sigma$, where $\sigma$ is the dead layer uncertainty, and evaluating the impact on the efficiency. Typical sizes of the detector dead layers are 1--2\,mm known with a typical uncertainty of 5--30\,\%~\cite{GERDA:2019vry}. 
The corresponding systematic uncertainty on the $\gamma$ detection efficiency is 3\% for \bege detectors, 4\% for \coax detectors, and 1\% for \ic detectors. The $\gamma$ detection efficiencies multiplied by the mass of LAr in the simulation volume, together with their uncertainties, are summarized in table~\ref{tab:res-eff} for the different data sets.

The two X-rays that are emitted in the process being searched for are neglected in the simulations. 
As anticipated in section~\ref{sec:exp-data}, their energy deposition in LAr could trigger the LAr veto. To account for this possibility, 
the survival probability of the two X-rays to the LAr veto cut is evaluated and combined with the $\gamma$ detection efficiency. 
We use the \gerda photon detection probability map developed in~\cite{GERDA:2022hxs} to estimate the probability $p(x,y,z)$ to detect 
scintillation light for each simulated event starting at position $(x,y,z)$ and corresponding to a full $\gamma$ energy deposition. 
From this probability, the number of photons $n$ produced by the two X-rays of total energy  $\text{E}_\text{X-rays}$ = (2.47 + 0.23)\,keV 
is obtained:
\begin{linenomath*}
\begin{equation}
\begin{aligned}
    n &= \text{E}_\text{X-rays} \cdot 28.12 \, \frac{\text{photons}}{\text{keV}} \cdot p(x,y,z) \;, 
\end{aligned}
\end{equation}
\end{linenomath*}
where 28.12 is the number of photons produced for an energy deposition of 1\,keV expected in the \gerda LAr~\cite{GERDA:2022hxs}. The probability P that the corresponding event survives the LAr veto cut is the Poisson probability $\mathcal{P}(0,n)$.\footnote{Where the probability mass function for a Poisson variable is defined as $\mathcal{P}(k,\lambda)=\frac{\lambda^k e^{-\lambda}}{k!}$.} The mean survival probability is obtained by averaging the survival probabilities of the events corresponding to a full $\gamma$ energy deposition and results in $\overline{P}=0.957$. Thus, the data selection discards almost 5\% of the events due to the X-rays depositing their energy in LAr. 
The calculation of the survival probability assumes that the two X-rays deposit all the energy at the exact point where the $\gamma$ ray is emitted. This assumption is considered valid since the attenuation length for a 3\,keV X-ray was estimated to be about 42\,$\mu$m~\cite{XrayCoefficients}, negligible compared to the $3\times 3\times 3$\,mm$^3$ binning of the photon detection probability map. The main systematic uncertainty on the mean survival probability comes from the photon detection probability map. The uncertainties on this probability map given in~\cite{GERDA:2022hxs} result in a 0.5\% systematic uncertainty on the survival probability.
Finally, we should note that the photon detection probability map assumes the pre-upgrade configuration of the LAr instrumentation~\cite{GERDA:2022hxs}. This means the model does not include the inner fiber shroud installed during the upgrade to improve the light detection efficiency near the germanium detectors~\cite{GERDA:2020xhi}. Therefore, a customized LAr veto cut was applied to select the post-upgrade data used in this work: the SiPM channels corresponding to the inner fiber shroud are not considered to build the LAr veto condition. This way, the X-rays survival probability obtained with the pre-upgrade photon detection probability map is extended to the post-upgrade data sets.

\section{Analysis methods}\label{sec:analysis}

The energy region used to set a limit on the half-life of \onecec of \Ar is defined between 410 and 450\,keV ($\pm$20\,keV around the $\gamma$ energy of 429.88\,keV, as indicated by the orange band in figure~\ref{fig:ar36_lowenergy_spectrum}). Given the high statistics in this energy region, data are used in a binned form, with a 1\,keV binning. It was checked that the binning choice did not impact the analysis results. In this energy region, the dominant backgrounds are the $\beta$ decay of $^{39}$Ar and the \nnbb decay of \Ge. Subdominant contributions to the background are, in order of importance, the $^{42}$K decays in LAr, the $^{40}$K, $^{214}$Pb, and $^{214}$Bi decays in structural materials. The sum of these contributions in the analysis window can be approximated by a linear distribution, as seen in figure~\ref{fig:ar36_lowenergy_spectrum}. The signal is modeled with a Gaussian peak centered at the $\gamma$ energy and with the width given by the detector energy resolution ($\sigma$ = FWHM/2.355). Uncertainties on the energy scale are parametrized by a shift of the signal peak $\delta$ compared to the nominal energy. 

A simultaneous fit is performed on the eight data sets listed in table~\ref{tab:res-eff} 
by adopting the following binned likelihood:
\begin{linenomath*}
\begin{equation}\label{eq:likelihood}
    \mathcal{L}(T_{1/2},\Vec{\theta}) = \prod_d \prod_i \mathcal{P}(n_{di}|\mu_{di}(T_{1/2},\Vec{\theta}_d)) \times \text{Pull}(\Vec{\theta}_d) \,,
\end{equation}
\end{linenomath*}
where the number of events in each bin is Poisson distributed, and the likelihood is given by the product of the Poisson probabilities $\mathcal{P}$ for all bins $i$ and data sets $d$. The likelihood depends on the half-life \HL of the investigated process, which is a common parameter among the eight data sets and is the only parameter of interest, and on some nuisance parameters $\Vec{\theta}$ that are data set specific and affect both the signal and background distributions. Gaussian pull terms $\text{Pull}(\Vec{\theta})$ are introduced in the likelihood to constrain some of the nuisance parameters. Finally, $n_{di}$ denotes the number of observed events in the data set $d$ and bin $i$, and $\mu_{di}$ is the expectation value for the same data set and bin. The latter is given by the sum of the signal and background in that bin: $\mu_{di}=b_{di}+s_{di}$. 
The number of signal events $s_{di}$ is given by the integral of the signal distribution for the data set $d$ in the bin $i$. This is a Gaussian distribution centered at $E+\delta_d(E)$, where $E$ is the $\gamma$ energy of 429.88\,keV and $\delta_d(E)$ the energy bias for the data set $d$ calculated for the same energy, and with the width given by the detector energy resolution $\sigma_d(E)=\text{FWHM}(E)/2.355$ evaluated for the same data set and at the same energy. The total number of signal events in a data set $d$ is related to the half-life \HL through the relation:
\begin{linenomath*}
\begin{equation}
    S_d = \ln(2)\cdot N_A \cdot \frac{m_{LAr,d}}{M_{36}}\cdot f_{36} \cdot t_d \cdot \varepsilon_{tot,d} \cdot \frac{1}{T_{1/2}} \,,
\end{equation}
\end{linenomath*}
where $N_A$ is the Avogadro constant, $M_{36}$ is the molar mass of argon (35.968\,g/mol), $m_{LAr,d}$ is the mass of LAr in the simulations from which the $\gamma$ detection efficiencies are extracted (the product $\varepsilon_\gamma \cdot m_{LAr}$ is given in table~\ref{tab:res-eff} for each analysis data set), $f_{36}$ is the abundance of \Ar in ultra-pure natural Argon (0.334\%)~\cite{lee2006}, and $t_d$ is the live time of the experiment. The total efficiency $\varepsilon_{tot,d}$ for the \PII data sets is given by the product $\varepsilon_{tot} = \varepsilon_\gamma \cdot \varepsilon_X \cdot \varepsilon_{LAr}$, where $\varepsilon_\gamma$ is the $\gamma$ detection efficiency, $\varepsilon_X$ the X-rays survival probability (both discussed in section~\ref{sec:efficiency}), and $\varepsilon_{LAr}$ is the efficiency of the LAr veto cut. The latter was estimated to be (97.7$\pm$0.1)\% for the pre-upgrade data and (98.2$\pm$0.1)\% for the post-upgrade data~\cite{GERDA:2020xhi}. The total efficiency of the \PI data sets equals the $\gamma$ detection efficiency $\varepsilon_\gamma$, because no LAr veto cut was available in \gerda \PI. 
Analogously, the number of background events $b_{di}$ is given by the integral of the background distribution for the data set $d$ in the bin $i$. The background distribution is a linear function that depends on two parameters, the normalization and the slope, both data set-specific. We verified that the first-order polynomial function describes the data in this energy region well and that a second-order polynomial function does not fit the data better. In modeling the background of \PI data, an additional Gaussian distribution is used to describe the full energy deposition of the 433.9\,keV $\gamma$ ray from $^{108m}$Ag, which lies in the energy region of the analysis. Contamination from $^{108m}$Ag was observed in the screening measurements, and all the three expected $\gamma$ lines from $^{108m}$Ag were observed in \gerda \PI data~\cite{Agostini2016,GERDA:2015naf}. The origin of the $^{108m}$Ag contamination in \gerda \PI was found in the signal cables~\cite{GERDA:2015naf}, which were exchanged in \gerda \PII~\cite{GERDA:2017ihb}. In addition, none of these $\gamma$ lines was observed in \gerda \PII data after the LAr veto cut. The decay of $^{108m}$Ag proceeds through a cascade of three equally probably $\gamma$ rays at energies of 433.9\,keV, 614.3\,keV, and 722.9\,keV. Therefore, even if any $^{108m}$Ag contamination were still present in \gerda \PII, the LAr veto cut would likely discard the corresponding events. 
In total, the fit has 42 floating parameters, 22 describing the signal peak ($\varepsilon_d$, $\delta_d$, $\sigma_d$)\footnote{While for the \PII data, the energy bias $\delta_d$ is assumed to be different among data sets, for \PI data only one parameter, common for the three data sets is adopted, following the previous analysis~\cite{Agostini2016}.}, 10 for the linear background of \PII data sets, 6 for the linear background of \PI data sets, 3 parameters for the number of $^{108m}$Ag events in \PI data sets, plus one common parameter to all data sets \invHL. The latter is constrained to positive values. 
Gaussian pull terms in the likelihood given in Eq.~\ref{eq:likelihood} constrain some of the nuisance parameters, namely the efficiency $\varepsilon_d$, the energy bias $\delta_d$, and the energy resolution $\sigma_d$ around their central value and uncertainty. All the other nuisance parameters are free and unconstrained, and their uncertainties are propagated into the result by profiling. 

To set a lower limit on the half-life of the investigated process, we use a modified frequentist approach, namely the \cls method~\cite{Read2002}. The latter was found to be a more appropriate choice in the case of an experiment with low sensitivity or, in different words, a background-dominated experiment~\cite{Read2002}. Compared to a pure frequentist approach, the \cls exclusion region does not assure the correct coverage and often results in an over-coverage, thus a more conservative result. The profile likelihood ratio test statistic is used for the \pvalue calculation. Asymptotic distributions of the test statistic and the Asimov data set are used~\cite{Cowan2011}. The statistics in each bin is high enough for this assumption to be valid. 

\section{Results}\label{sec:result}

\begin{figure*}
\centering
  \includegraphics[width=\textwidth]{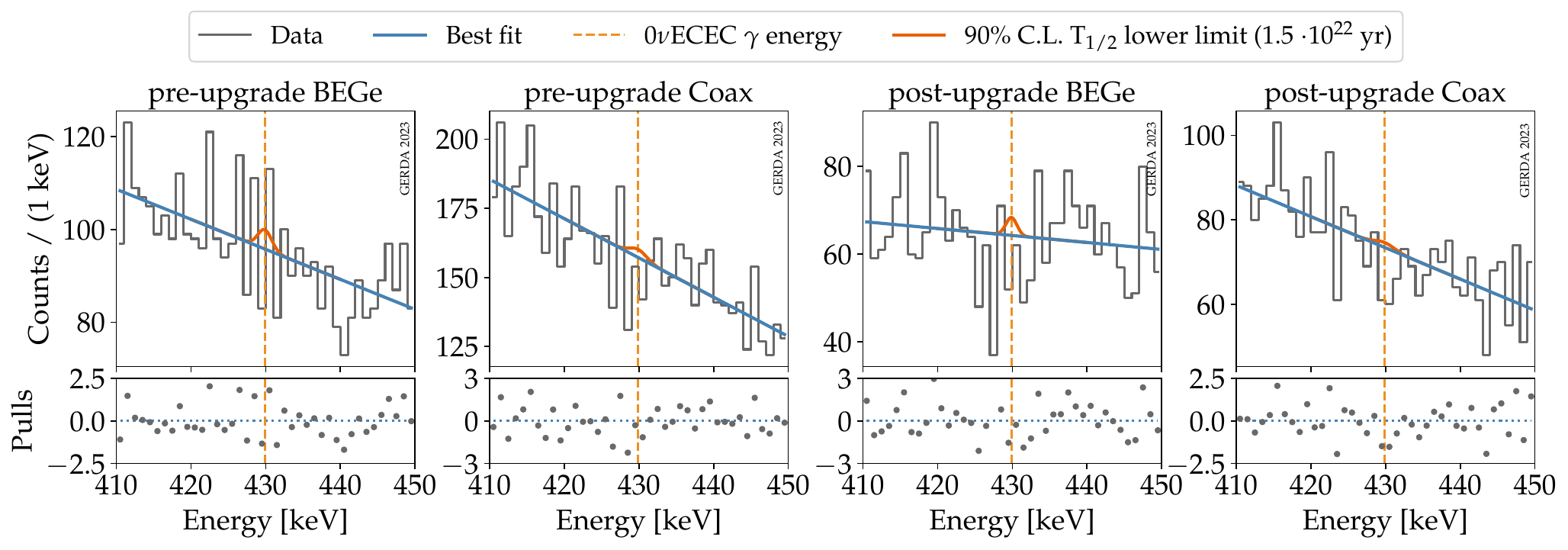} \\
  \includegraphics[width=\textwidth]{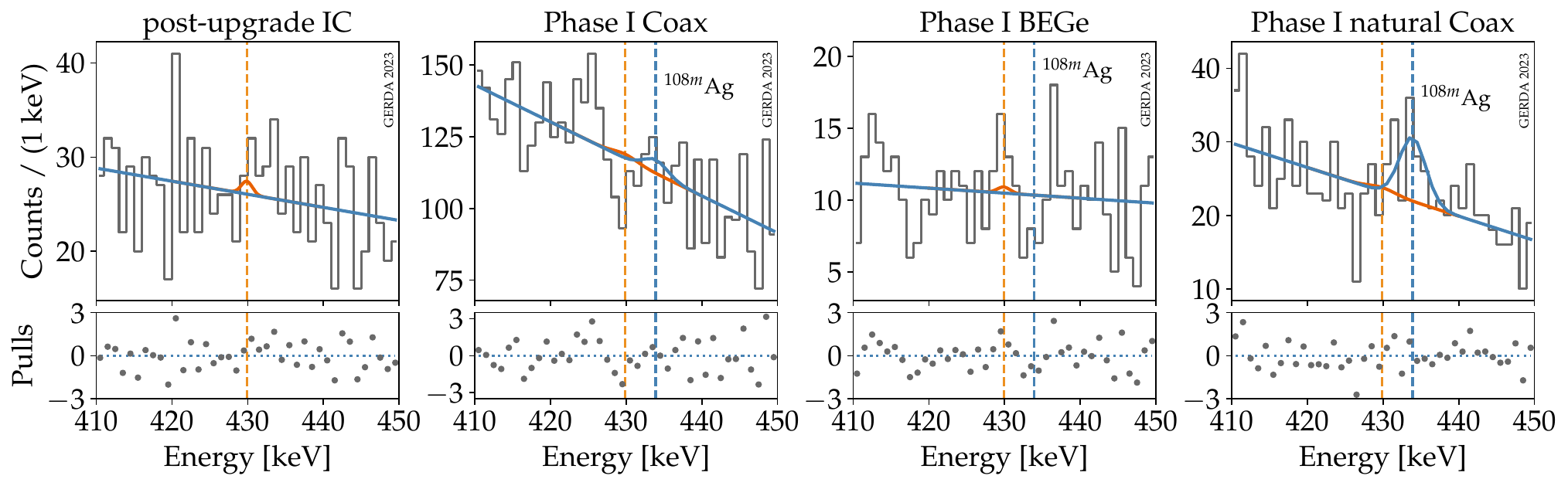}
\caption{Best fit of the combined \gerda data. The blue line shows the combined best-fit model, corresponding to \invHL= 0. The dashed orange line indicates the energy at which a $\gamma$ line from \onecec is expected, and the orange peak displays the expected signal for a half-life equal to the 90\% C.L. lower limit $1.5\cdot 10^{22}$\,yr. The pulls, {\it i.e.} residuals normalized to the expected statistical fluctuations of the bins, are shown in the bottom panels for each data set. }
\label{fig:fit-result-data}   
\end{figure*}

\begin{figure}
  \includegraphics[width=\columnwidth]{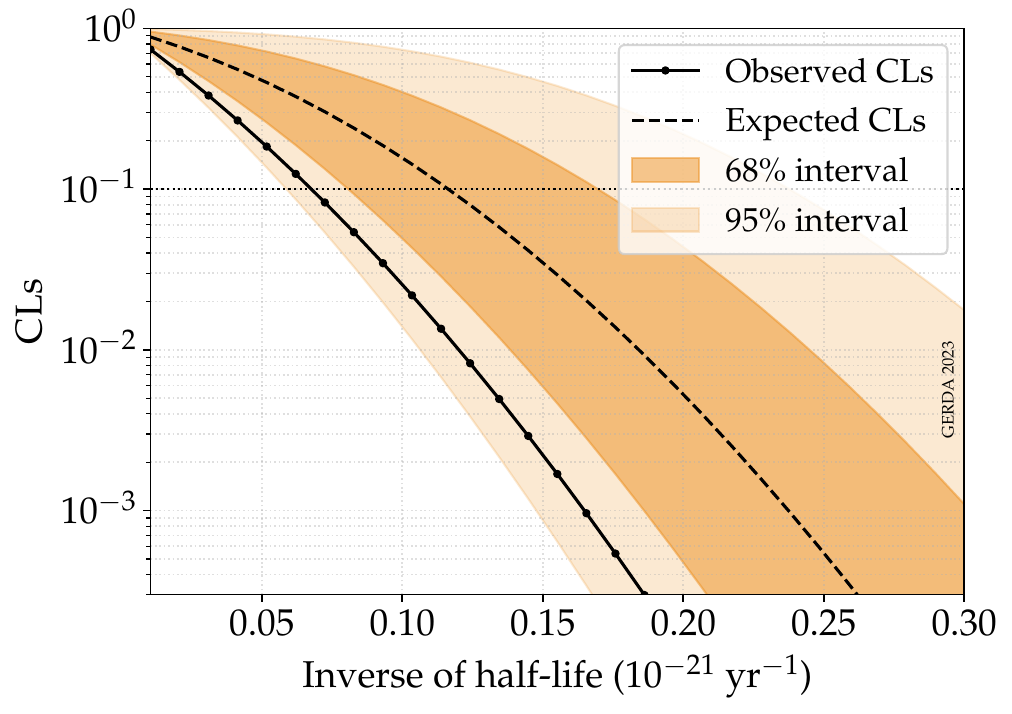}
\caption{\cls as a function of the inverse of the half-life obtained in the analysis of the combined \gerda \PI and \PII data. The median of the \cls distribution for the \gerda experiment under the no signal hypothesis and the observed \cls for the \gerda data are shown by the continuous black line and the dashed line, respectively. The spread of the \cls expected distribution, given by the 68\% and 95\% probability intervals, is also shown by the colored bands. The 90\% C.L. limit (sensitivity) is given by the solid (dashed) black line intersection with the dotted line, corresponding to a \cls of 0.1.}
\label{fig:CLs-scan}      
\end{figure}

The best fit, defined as the minimum of the profiled likelihood ratio, yields \invHL= 0, {\it i.e.} we do not observe any signal events from \onecec. Data from the five \gerda \PII analysis data sets and in the energy region of the analysis are shown in figure~\ref{fig:fit-result-data} together with the best-fit model and the residuals normalized to the expected statistical fluctuations of the bins. 
The 90\% C.L. limit on the half-life is obtained by scanning the observed \cls over different values of \invHL and finding the value for which \cls= 0.1. For \gerda \PII data only, this gives $T_{1/2}>1.3\cdot 10^{22}$\,yr. The 90\% C.L. sensitivity of the \gerda \PII experiment, {\it i.e.} the median expectation under the no signal hypothesis, is obtained analogously by scanning the expected \cls over different values of \invHL and finding the value for which \cls= 0.1. The latter gives $T_{1/2}>8.0\cdot10^{21}$\,yr. The analysis of the combined \gerda \PI and \PII data gives a 90\% C.L. sensitivity of $T_{1/2}>8.6\cdot 10^{21}$\,yr and an observed lower limit of $T_{1/2}>1.5\cdot 10^{22}$\,yr. Figure~\ref{fig:CLs-scan} shows the scan of the observed and expected \cls over a range of values of \invHL obtained in the analysis of the combined \gerda \PI and \PII data.

Systematic uncertainties on the efficiency $\varepsilon_d$, the energy bias $\delta_d$, and the energy resolution $\sigma_d$ are identified as primary sources of systematic uncertainties and included in the likelihood through nuisance parameters constrained by Gaussian pull terms as explained in section~\ref{sec:analysis}. Their overall effect on the limit derived in section~\ref{sec:result} is estimated to be 2\%. Potential systematic uncertainties related to the fit model, particularly the background distribution, are also investigated. First, the assumption of a linear distribution is compared to a more general second-order polynomial distribution. This has a negligible impact on the result. The presence of additional structures in the background is also investigated. As discussed in section~\ref{sec:analysis}, a $\gamma$ line from $^{108m}$Ag, very close to the expected signal energy, is included in the background model of the \PI data sets, as in previous analysis~\cite{Agostini2016}. A possible systematic uncertainty due to the above $\gamma$ line in \PII data is investigated by introducing it in the background model. This would worsen our result of a 2\%.

\section{Conclusions}
In this work, we searched for the 429.88\,keV $\gamma$ line from the \Ar \onecec using the final total exposure of the \gerda \PII experiment, combined with the \gerda \PI exposure. No signal was observed, and a lower limit on the half-life of this process was derived, yielding $T_{1/2}>1.5\cdot 10^{22}$\,yr (90\% C.L.). This is the most stringent limit on the half-life of the \Ar \onecec. This work shows that the potential of the \gerda experiment in investigating physics beyond the Standard Model extends further than the search for the \onbb decay of \Ge (see also~\cite{GERDA:2020emj,GERDA:2022ffe}). Even if the sensitivity is many orders of magnitude below the theoretical expectation for this process, to our knowledge, the \gerda experiment was, to date, the only experiment with the capability to search for the \onecec of \Ar with competitive sensitivities. 
The \gerda sensitivity is limited by the physical background from $^{39}$Ar $\beta$ and \Ge \nnbb decays in the energy region where the $\gamma$ peak is expected, which is, for instance, orders of magnitude higher than the background in the region of interest for the \Ge \onbb decay.  An additional limiting factor is the low detection efficiency since the $\gamma$ ray is emitted in the LAr and must be detected in one of the germanium detectors. Only $\gamma$ rays emitted in the proximity of the detector array contribute to the total efficiency as discussed in section~\ref{sec:efficiency} (See figure~\ref{fig:eff_sim}). 

Among the planned future experiments, the Large Enriched Germanium Experiment for Neutrinoless-$\beta\beta$ Decay (LEGEND) experiment can extend the search for the \onecec of \Ar to higher sensitivity. In the first phase of the project, LEGEND-200 will deploy about 200\,kg of germanium detectors. This is more than a factor of four compared to the \gerda detector mass and will imply a higher detection efficiency to the $\gamma$ ray emitted in this process. On the other hand, the background in the energy region where the $\gamma$ peak is expected should be comparable to the \gerda background, largely dominated by the $^{39}$Ar $\beta$ decay. Still, an improvement in the current sensitivity is foreseen. LEGEND-1000 will deploy about 1 ton of germanium detectors, implying an even higher detection efficiency to the $\gamma$ ray emitted in this process. In addition, using underground Ar instead of atmospheric Ar is intended. This is depleted of $^{39}$Ar, which is the main background contribution in this search. A significant improvement in the sensitivity is therefore expected. To our knowledge, no other planned experiment has competitive sensitivity to LEGEND in the search for \onecec of \Ar.


\begin{acknowledgements}
The \textsc{Gerda} experiment is supported financially by
the German Federal Ministry for Education and Research (BMBF),
the German Research Foundation (DFG),
the Italian Istituto Nazionale di Fisica Nucleare (INFN),
the Max Planck Society (MPG),
the Polish National Science Centre (NCN, grant number UMO-2020/37/B/ST2/03905),
the Polish Ministry of Science and Higher Education (MNiSW, grant number DIR/WK/2018/08),
the Russian Foundation for Basic Research,
and the Swiss National Science Foundation (SNF).
This project has received funding/support from the European Union's
\textsc{Horizon 2020} research and innovation programme under
the Marie Sklodowska-Curie grant agreements No 690575 and No 674896.
This work was supported by the Science and Technology Facilities Council, part of the U.K. Research and Innovation (Grant No. ST/T004169/1).
The institutions acknowledge also internal financial support.
 
The \textsc{Gerda} collaboration thanks the directors and the staff of
the LNGS for their continuous strong support of the Gerda experiment.
\end{acknowledgements}

This manuscript has associated data in a data repository. [Authors’ comment: The data shown in Figs.~\ref{fig:ar36_lowenergy_spectrum} and~\ref{fig:fit-result-data} is available in ASCII format as Supplemental Material~\cite{supp-mat}.]

\bibliographystyle{spphys}       
%

\end{document}